\documentclass[prb,superscriptaddress,showpacs,twocolumn]{revtex4}

\usepackage{graphicx}

\usepackage{amssymb,amsmath}
\usepackage{subfigure}

\usepackage[dvips]{color}
 
\newcommand{\eref}[1]{Eq.~(\ref{#1})}
\newcommand{\fref}[1]{Fig.~\ref{#1}}

\newcommand{\tref}[1]{Tab.~\ref{#1}}

\newcommand{\im}{%
           \imath}
\newcommand{\bra}[1]{\ensuremath{\langle #1|}}
\newcommand{\ket}[1]{\ensuremath{|#1\rangle}}

\def\egp{\ensuremath{e_g}}

\def\t2g{\ensuremath{t_{2g}}}
\def\a1g{\ensuremath{a_{1g}}}

\newcommand{\svek}{%
        \mathbf}

\newcommand{\vek}[1]{%
        \hbox{\textbf #1}}

\newcommand{\pr}{%
        ^\prime}

\def\tio2{TiO\ensuremath{_2}}

\def\sio2{SiO\ensuremath{_2}}
\def\etal{{\it et~al.}}

\def\XXint#1#2#3{{\setbox0=\hbox{$#1{#2#3}{\int}$}
\vcenter{\hbox{$#2#3$}}\kern-.5\wd0}}

\begin{document}

\author{ Jan M. Tomczak}
\affiliation{Research Institute for Computational Sciences, AIST, Tsukuba, 305-8568 Japan}
\affiliation{Japan Science and Technology Agency, CREST}
\author{ T. Miyake}
\affiliation{Research Institute for Computational Sciences, AIST, Tsukuba, 305-8568 Japan}
\affiliation{Japan Science and Technology Agency, CREST}
\author{ F. Aryasetiawan}
\affiliation{Graduate School of Advanced Integration Science, Chiba University, Chiba, 263-8522 Japan}
\affiliation{Research Institute for Computational Sciences, AIST, Tsukuba, 305-8568 Japan}
\affiliation{Japan Science and Technology Agency, CREST}

\title{Realistic many-body models for Manganese Monoxide under pressure}

\begin{abstract}
In materials like transition metals oxides where electronic Coulomb correlations
impede a description in terms of standard band-theories, the application of genuine
many-body techniques is inevitable.
Interfacing the realism of density-functional based methods with the virtues of Hubbard-like Hamiltonians, requires the joint {\it ab initio} construction of transfer integrals and interaction matrix elements (like the Hubbard U) in a localized basis set.
In this work, we employ the scheme of maximally localized Wannier functions and the
constrained random phase approximation to create effective low-energy models
for Manganese monoxide, and track their evolution under external pressure.
We find that in the low pressure antiferromagnetic phase, the compression results in an {\it increase} of the bare Coulomb interaction for specific orbitals. 
As we rationalized in recent model considerations $[Phys. Rev. B 79, 235133 (2009)]$, this seemingly counter-intuitive behavior is a consequence of the delocalization of the respective Wannier functions. The change of screening processes  does not alter this tendency, and thus, the 
screened on-site component of the interaction -- the Hubbard $U$ of the effective low-energy system -- increases with pressure as well. The orbital anisotropy of the effects originates from the orientation of the orbitals vis-\`a-vis the
deformation of the unit-cell. 
Within the high pressure paramagnetic phase, on the other hand,
we find the significant increase of the Hubbard $U$ is insensitive to the orbital orientation and almost exclusively owing to
a substantial weakening of screening channels upon compression.
\end{abstract}

\maketitle

\section{Introduction}
\subsection{Realistic models for correlated materials}
While standard band structure methods, like density functional theory (DFT) in the local spin density approximation (LSDA),\cite{RevModPhys.71.1253,RevModPhys.61.689}
in principle allow for a treatment of the realistic complexity of materials arising from orbital, spin and structural degrees of freedom,
the failure of these techniques in the presence of strong electronic correlations is well documented.\cite{vollkot}
Many-body techniques, on the other hand, while potentially treating correlation effects with accuracy, often lack the realism to
account for the electronic structure of a material from a first-principles basis.

Hence, when it comes to the {\it ab initio} description of correlated matter, the joining of these two fields is of paramount importance.
Often, this combining amounts to constructing effective, yet realistic, low energy models, in the spirit of the Hubbard or Anderson model.
These consist of a one-particle part, extracted from the Kohn-Sham band structure, and two-particle interaction terms.
There are different paradigms for choosing the matrix elements of the latter.
Sometimes, these values are treated as mere empirical parameters, adjusted to correctly reproduce some experimental finding.
However, there exist first-principle methods, among which the most popular are
the constrained LDA
technique,\cite{constrainedLDA} and the constrained random phase approximation (cRPA).\cite{PhysRevB.70.195104}

In this article we are chiefly concerned with the evolution of interaction matrix elements under pressure within a realistic setup.
While the influence of pressure onto structures and band-structures has been widely studied on the DFT level, changes in the interaction have received
much less attention. Yet, the pronounced sensitivity of correlated matter on external parameters~\cite{imada} heralds the need for an accurate treatment
of all ingredients to the electronic structure. For the LDA+U~\cite{PhysRevB.44.943} technique, this issue has been investigated within
a linear response based approach~\cite{PhysRevB.71.035105,tsuchiya:198501,hsu:125124}, and the reproduction of crystal and electronic structures, as well as phase stabilities were indeed found to require the description of pressure induced changes in the Hubbard $U$.

Here, we are interested in setting the stage for many-body methods that go beyond density functional methods.
In this vein, a particularly versatile technique 
is the use of maximally localized Wannier functions~\cite{PhysRevB.56.12847,PhysRevB.65.035109} within the cRPA setup,\cite{miyake:085122,jmt_wannier} which is the method of choice of this article.
In this formalism it is possible to turn off precisely those screening channels that
are to be left to the solution of the effective model.\cite{PhysRevB.70.195104}$^,$\footnote{In the constrained LDA technique for the Hubbard $U$ of e.g.\ a d-manifold, the screening originating from transitions of other bands to and from the d-bands are eliminated, although they are not accountable for by the low energy system -- a reason why the constraint LDA technique often overestimates the values of U.}
Moreover, the use of many-body techniques often requires the use of localized basis sets, and the Wannier construction is geared at exactly this.
While the Wannier-cRPA technique has been already applied to several systems, such as 3d transition metals,\cite{miyake:085122} the perovskite SrVO$_3$,\cite{miyake:085122} and oxypnictides~\cite{JPSJS.77SC.99,JPSJS.77.093711},  the evolution of the respective Coulomb interaction matrix elements under external stimuli such as pressure have not been addressed in the realistic case.
In recent model considerations,\cite{jmt_wannier} we established generic behaviors of maximally localized Wannier functions and interaction matrix elements under pressure for simple setups. We rationalized that as a consequence of the delocalization of Wannier functions under compression,
not only do transfer integrals increase, but also -- counterintuitively -- the matrix elements of the bare Coulomb interaction augment.
In this paper, we extent this work to a more complex scenario by
applying the fully {\it ab initio} Wannier-cRPA formalism
to manganese monoxide, MnO, under pressure.
In particular, we monitor the evolution of transfer integrals, on-site matrix elements of the bare Coulomb interaction, and of the partially screened Coulomb interaction (the Hubbard U, and the Hund J), 
for different realistic low-energy systems. We will distinguish between pressure induced changes related to the construction of the sub-system, and changes in the screening strength as provided by the band-structure and transition matrix elements in the polarization. 
We find that while the former mainly determine the behavior of the interactions in the low temperature, low pressure anti-ferromagnetic phase of MnO, changes in the Hubbard $U$ of the high pressure paramagnetic phase are dominated by screening related effects.
\subsection{Manganese monoxide}
\subsubsection{Approaches to the crystal and electronic structure}
Concerning the failures of the LDA in the realm of correlated materials, the focus material of this work, manganese monoxide, MnO, is rather well behaved at first glance~: 
Below 118~K the system is antiferromagnetic and due to the fact that the Mn$3d$ orbitals are half-filled, 
spin-dependent band-structure  calculations~\cite{PhysRevB.30.4734,PhysRevB.59.762,Cohen01311997,franchini:045132,franchini:195128,Zhang_mno,kasinathan:195110,tran:155108} do indeed yield an insulator.
However, above the N\'eel temperature, which is much smaller than the experimental charge gap of 3.9~eV,\cite{PhysRevB.44.1530} MnO is paramagnetic,
and believed to be a Mott insulator.
A straight forward band-structure calculation in the paramagnetic phase, however, yields a metal.

A genuine feature of correlated materials is their sensitivity to external stimuli, leading to rich phase-diagrams with regimes of different spin or orbital orders, and metal-insulator transitions.\cite{imada}
MnO is no exception and both experiments and theory helped to elucidate the phases of manganese monoxide as a function of temperature versus pressure. See Fig. 1 in Ref.~\cite{yoo:115502} for a graphical depiction of compiled results.

The paramagnetic insulator MnO, which is of rock-salt structure (B1), 
contracts below the N\'eel temperature along the $\langle 111\rangle$ direction, resulting in a rhombohedral distorted phase (dB1).
MnO has been the prototype material for the discovery and description of the superexchange mechanism.\cite{Kramers1934182,PhysRev.79.350,mno_zs}
It was rationalized by Terakura \etal~\cite{PhysRevB.30.4734} that this strong manganese superexchange via the oxygen orbitals~\cite{Kramers1934182,PhysRev.79.350} is responsible for a spin oder that is of 
AF II type,\cite{PhysRev.76.1256.2, PhysRev.83.333,PhysRevB.38.11901} with alternating spins along the distortion axis.
With increasing pressure, the N\'eel temperature rises~\cite{Webb,Bloch_mno} and reaches room temperature (300~K) at 30GPa.\cite{yoo:115502}
It has been shown that beyond the volume compression, it is the rhombohedral distortion that stabilizes the AF II order under pressure with respect to other spin orientations.\cite{PhysRevB.59.762}
At constant temperature (300~K), a further transition occurs at 90~GPa,\cite{Syono_mno,kondo:4153,yoo:115502} when 
the system becomes, again, a paramagnetic insulator, yet of NiAs (B8) structure.\cite{yoo:115502}
Notably, the high pressure B8 structure was predicted by DFT calculations in the generalized gradient approximation (GGA).\cite{PhysRevB.59.762}
At still higher pressure, the system undergoes an iso-structural Mott insulator to metal transition at 105~GPa,\cite{yoo:115502}
signaled by a volume collapse of 6.6\%,\cite{yoo:115502} metallic lustre,\cite{kondo:4153} an increase of reflectance,\cite{Kobayashi01,mita:100101} a resistance decrease
of several orders of magnitude~\cite{PhysRevB.69.220101} and a quenching of the magnetic moment.\cite{yoo:115502}
The transition was anticipated, from theoretical considerations~\cite{Ohnishi} and {\it ab initio} calculations.\cite{Cohen01311997}
Yet, both works forward different mechanism~: In Ref.~\cite{Ohnishi} a pressure induced increase in the crystal-field was invoked to explain the metalization, while in Ref.\cite{Cohen01311997}, in which the rhombohedral distortion is neglected, a growing band-width was seen at the origin of the transition.
We will come back to this in the discussion of our results, which favor the first point of view.

On the theory side,
numerous works that aim at reproducing the described phase-diagram from first principles resort to improvements within the
density functional formalism. Among these are the generalized gradient approximation (GGA),\cite{Cohen01311997,PhysRevB.59.762,kasinathan:195110,Zhang_mno,franchini:045132,franchini:195128,tran:155108}
self-consistent Hartree,\cite{PhysRevB.50.5041} LDA/GGA+U,\cite{PhysRevB.59.762,franchini:045132,franchini:195128,1367-2630-9-7-235,Zhang_mno,kasinathan:195110,1367-2630-9-7-235} self-interaction corrected (SIC) LSDA.\cite{PhysRevLett.65.1148,PhysRevB.47.4029}
Also applied were various hybrid functional~\cite{becke:1372} based approaches~\cite{tran:155108,kasinathan:195110,franchini:195128,franchini:045132}.

Although relying on the coincidence of a half-filled d-shell and the spin order for insulating results,
it is due to the fact that (unlike for the other transition metal monoxides) the $3d$ electrons are not falsely itinerant,
that the above methods succeed in yielding at times reasonable structures, lattice constants, bulk moduli,
magnetic moments, and sometimes even values for the charge gap (see especially the SIC and hybrid functional references from above).
Results have been obtained also within the GW approximation.\cite{hedin,ferdi_gw,RevModPhys.74.601}
While pioneering work~\cite{PhysRevLett.74.2323} still had to resort to numerical simplifications, the increase in computer performance
nowadays even allows the application of self-consistency schemes to MnO,\cite{PhysRevLett.93.126406}
which considerably improve on the gap value of the antiferromagnetic phase with respect to LDA calculations.

Yet, despite these successes, it is evidently desirable to have a many-body technique that works in a wider context, capturing correlated metals, ordered {\it and} Mott insulators.
In this vein, the probably most generally applicable approach nowadays is LDA+DMFT,\cite{RevModPhys.78.865}
where a Hubbard Hamiltonian consisting of a one-particle part stemming from the LDA and an interaction part of Hubbard-Hund type
is solved within dynamical mean field theory (DMFT).\cite{bible}$^,$%
\footnote{
This discards correlation induced changes in the charge density. See e.g.\ \cite{pourovskii-2007} for a charge self-consistency scheme within DMFT.
} 
While DFT methods and their generalizations, including LDA+U,\cite{PhysRevB.44.943} cannot address finite-temperature insulators without broken symmetries, LDA+DMFT is in principle capable to deal with all the different phases. 
Concerning MnO, the technique has first been used for the calculation of phonon dispersions.\cite{PhysRevLett.90.056401}
Recently LDA+DMFT was applied by Kune\v{s} \etal~\cite{kunes_mno} to study the isostructural Mott insulator to metal transition witnessed at 105GPa (300~K). The results reproduced quite accurately all experimental findings, in particular the critical pressure and the joint collapse of volume and magnetic moment.\cite{kunes_mno}

Often it is criticized that LDA+$U$ and LDA+DMFT calculations are not fully {\it ab initio} since the interactions 
are treated as ``parameters'' that are in a sense tuneable. 
Indeed in the above LDA+DMFT work,\cite{kunes_mno} the Hubbard $U$ and the Hund $J$ were chosen to be constant as a function of pressure.
Although adjustable parameters are to be preferred 
over artificially broken phases, it is true that especially in transition metal monoxides there are
several competing energy scales~\cite{PhysRevLett.55.418} so that the values of the interactions have to be known with greater precision than in other compounds.
Among those relevant energies are~: the crystal field splitting, that measures the 
degree of non-degeneracy of the centers of gravity of the d orbitals, e.g.\ $\Delta_{cf}=\bar{\epsilon}_{\egp}-\bar{\epsilon}_{\t2g}$ for an octahedral coordination, the charge transfer energy $\Delta_{ct}=\bar{\epsilon}_{\t2g}-\bar{\epsilon}_{2p}$
here defined between the O$2p$ and the Mn \t2g, the transfer integrals (hoppings) $t_{\t2g}$, $t_{\egp}$, the hybridizations $t_{pd}$, and the on-site Coulomb interaction $U$ of the d orbitals.
\subsubsection{Earlier works on the Hubbard U}
In the pioneering work of Anisimov~\etal~\cite{PhysRevB.44.943} that introduced the LDA+$U$ formalism, the constrained LDA technique was applied to MnO at zero pressure and a value $U=6.9$~eV was found. 
The Hubbard $U$ is thus of the same order as the charge transfer energy $\Delta_{ct}$ (see below). Therewith MnO is in the intermediate regime between a charge transfer and a Mott-Hubbard insulator~\cite{PhysRevLett.55.418} and changes in $U$ might considerably alter the many-body spectra and their interpretation.
 
An empirical technique that was applied to MnO is the fitting of
configuration interaction (CI) cluster calculations to photoemission experiments.
Although of course depending on the details of the chosen cluster, we cite the values
$U=7.5$~eV, $\Delta_{ct}=7.0$~eV of Fujimori \etal~\cite{PhysRevB.42.7580} and $U=8.5$~eV, $\Delta_{ct}=8.8$~eV of van Elp \etal.\cite{PhysRevB.44.1530}
Still, CI cluster calculations cannot account for delocalized bands. As a consequence, the struggle of d electrons between localized and itinerant behavior and the mixing with the more extended O$2p$ and Mn4s orbtials is biased. 

Using a semi-empirical Anderson (d-) impurity Hamiltonian, Zaanen and Sawatzky~\cite{mno_zs} estimated $\Delta$ and $U$ to be both around $9$~eV,
and found accurate values also for the N{\'e}el temperature.
An Anderson Hamiltonian was recently also applied to results from
resonant inelastic x-ray scattering (RIXS) spectra~\cite{ghiringhelli:035111}~: the parameters of a single manganese d impurity model were adjusted to yield $U=7.2$~eV and $\Delta_{ct}=6.5$~eV.
 In this approach, the delocalized character of the oxygen and Mn4s orbitals is better captured than within CI based methods.
Indeed, this is a step toward LDA+DMFT calculations, albeit without obeying the self-consistency condition that relates the manganese impurity with the periodic solid.

The Hubbard $U$ can also be extracted from GW calculations. Indeed the RPA technique is a crucial ingredient to the GW approximation. 
Pioneering model GW calculations suggested a value of $U=8.0$~eV~\cite{PhysRevLett.74.2323} for the antiferromagnetic phase, albeit only empirically extracted from the energy difference between 
occupied and unoccupied $e_g$ and \t2g states.

\section{Method}
The method of using the Wannier orbital construction in conjunction with the constrained RPA has been presented in detail in Ref.~\cite{miyake:085122}. 
Here, we shall just give a brief summary as is necessary for understanding our results.
\subsection{Band-structure}
In this work, we employ the LSDA in the full-potential (FP) LMTO~\cite{fplmto}
realization and use experimental structures and lattice parameters as provided by
Yoo~\etal~\cite{yoo:115502} at 300~K~\footnote{Still, both LDA and RPA work at $T=0$.} and variable pressure.
We use up to 4096 reducible Brillouin zone points, the LMTO basis set includes orbitals up to $l=4(3)$ for Mn(O), and we use local orbitals~\cite{fplmto}
to incorporate semi-core and high-lying states.
\subsection{Maximally localized Wannier functions}
The one-particle part of the effective low energy systems is extracted from the (LMTO) band-structure calculation by the construction
of a Wannier basis~\cite{RevModPhys.34.645} of the desired subset of orbitals. 
This procedure not being unique,\cite{RevModPhys.34.645} we choose to use Wannier functions that are maximally localized.\cite{PhysRevB.56.12847,PhysRevB.65.035109}
While the initial Hilbert space is spanned by all Kohn-Sham wave functions, $\mathcal{H}=span\{\psi^{\hbox{\tiny KS}}_{\svek{k}\alpha}\}$, for the aim of constructing effective low energy models necessitates the choosing of a subspace, $\mathcal{H}^{low} \subset \mathcal{H}$, made up by
only selective Kohn-Sham orbitals. The Wannier functions are then deduced from the latter. 
Complications occur if the bands of the orbitals that constitute the desired sub Hilbert space are entangled with high energy bands.
For details on how then to construct maximally localized Wannier functions see~\cite{PhysRevB.65.035109}, and
for a comparison with other Wannier function schemes, for instants the approach in which the on-site screened Coulomb interaction is maximized, see e.g.\ the Refs.~\cite{RevModPhys.35.457,miyake:085122,jmt_wannier}.
\subsection{constrained RPA}
The matrix elements of the bare Coulomb interaction $v(\svek{r},\svek{r}\pr)=\frac{e^2}{4\pi\epsilon_0}1/\left| \svek{r}-\svek{r}\pr\right|$ in the Wannier basis $\chi^{\hbox{\tiny W}*}_{\svek{R}\alpha}(\svek{r})$ are given
by
\begin{eqnarray}
		&&V^{\alpha\beta\alpha\pr\beta\pr}_{\svek{R},\svek{R}\pr}=\frac{e^2}{4\pi\epsilon_0} \,\times\\
		&&\quad\int d^3r d^3r\pr \chi^{\hbox{\tiny 
		W}*}_{\svek{R}\alpha}(\svek{r}) \chi^{\hbox{\tiny W}}_{\svek{R}\beta}(\svek{r})\frac{1}{\left|\vek{r}-\vek{r}\pr\right|}
	\chi^{\hbox{\tiny W}*}_{{\svek{R}\pr}\alpha\pr}(\svek{r}\pr) \chi^{\hbox{\tiny W}}_{\svek{R}\pr\beta\pr}(\svek{r}\pr)\nonumber 
	\label{eqV}
\end{eqnarray}
From the latter, partially screened interaction matrix elements are obtained using the cRPA technique.\cite{PhysRevB.70.195104,miyake:085122}
Using the Kohn-Sham orbitals, the total RPA polarization can be expressed by
\begin{eqnarray}
	&&P(\vek{r},\vek{r}\pr,\omega)=\\
	&&\quad\sum_{spin}\sum_{\svek{k}n}^{occ}\sum_{\svek{k}\pr n\pr}^{unocc} 
	\psi^{\hbox{\tiny KS}*}_{\svek{k}n}(\vek{r})\psi^{\hbox{\tiny KS}\phantom{*}}_{\svek{k}\pr n\pr}(\vek{r})\psi^{\hbox{\tiny KS}*}_{\svek{k}\pr n\pr}(\vek{r}\pr)\psi^{\hbox{\tiny KS}\phantom{*}}_{\svek{k}n}(\vek{r}\pr) \nonumber\\
	&&\times
	 \left\{ \frac{1}{\omega-\epsilon_{\svek{k}\pr n\pr}+\epsilon_{\svek{k} n} +\im 0^+} - \frac{1}{\omega+\epsilon_{\svek{k}\pr n\pr}-\epsilon_{\svek{k} n} - \im 0^+}    \right\}\nonumber
	\label{pol}
	\end{eqnarray}
Its matrix elements within the Wannier basis set are given analogous to the above bare interaction.
The strength of screening channels can thus be influenced by the overlap integrals (matrix elements) of the wave functions, and by the energy difference of the Kohn-Sham excitations that appear in the denominators.

From the form of the fully screened interaction, $W(\omega )=\left[1-P(\omega )v\right]^{-1}v$ (that also appears in the GW formalism~\cite{hedin,ferdi_gw,RevModPhys.74.601})
it can be shown,\cite{PhysRevB.70.195104} that the screening contributions are actually additive.
Indeed one can split the polarization, $P=P_{low}+P_r$, into the transitions within a low-energy orbital subspace, $P_{low}$ of $\mathcal{H}^{low}$, and the rest $P_r=P_{high-high}+P_{high-low}+P_{low-high}$,
where the latter includes, both, transitions within the high-energy subspace, and between the two subspaces.
Then the fully screened interaction can be written recursively, i.e.\ 
$	W={\left[1-P_{low}W_r^{low}\right]^{-1}}{W_r^{low}}$
where
\begin{eqnarray}
	W_r^{low}&=&\frac{v}{1-P_{r}v}
\end{eqnarray}
is the Coulomb interaction when all screening processes that do not involve transitions within the low-energy orbital subspace have been accounted for%
\footnote{
In case the bands of the "low" and "high" subspace are entangled, the separation $P=P_{low}+P_r$ is not uniquely defined.
For a way to overcome this arbitrariness see the recent \cite{miyake:155134}.
Since in our case the entanglement is not too pronounced we do not employ this refinement.
}
.

\begin{table}%
\begin{tabular}{c||c|c}
setup & low energy subspace $\mathcal{H}^{low}$ & polarization $P_{low}$\\
\hline
\hline
d & $span\left\{ \psi^{KS}_{\svek{k}\alpha} \left|\right.  \alpha=\hbox{Mn}3d \right\}$ & $P_d$\\
pd & $span\left\{ \psi^{KS}_{\svek{k}\alpha} \left|\right.  \alpha=\hbox{Mn$3d$, O$2p$}\right\}$ & $P_d$$+$$P_p$$+$$P_{pd}$$+$$P_{dp}$\\
pd+p &$span\left\{ \psi^{KS}_{\svek{k}\alpha} \left|\right.  \alpha=\hbox{Mn$3d$, O$2p$}\right\}$&  $P_d$\\
\end{tabular}
\caption{Summary of the low-energy setups that are used~: Shown are the choices for the construction of the low energy subspace, and the polarization. See text for details.}
\label{tmodel}
\end{table}

For our compound, MnO, we will consider three different setups, that are also summarized in \tref{tmodel}.
\begin{itemize}
	\item In the ``d-setup'', a many body model is obtained from maximally localized Wannier functions that are constructed for the space of the Mn$3d$ Kohn-Sham (KS) orbitals, $\psi^{KS}_{\svek{k},3d}$, and from a partially screened interaction $W_r^{d}$ that is given by constraining the polarization to 
transitions that are not fully within the d-subspace%
\footnote{
While the RPA screening and thus also the effective interactions are energy dependent, so far, most
many-body techniques only use static interactions, which is why we will discuss in this work
only the low energy limit, i.e. for example the Hubbard $U_{\alpha\beta}={W_r}_{\svek{0},\svek{0}}^{\alpha\alpha,\beta\beta}(\omega=0)$ (using the conventions of \eref{eqV}). For a discussion on the frequency dependence of the
screened interaction see the Ref.~\cite{PhysRevB.70.195104}.
}$^,$
\footnote{
On a more general note, we stress, that the Wannier cRPA approach can be viewed as a first step towards the already mentioned combination of GW~\cite{hedin,ferdi_gw,RevModPhys.74.601} with the dynamical mean field theory (DMFT)~\cite{bible} ``GW+DMFT'',\cite{PhysRevLett.90.086402} in which the Hubbard $U$ is promoted from a parameter to a quantity that obeys a self-consistency relation.
}%
. 
\item MnO being close to the charge transfer regime, it is sensible to include the oxygen $2p$ orbitals for a realistic many-body treatment.
In such a ``pd-setup'' one constructs a Wannier basis for the space of Mn$3d$ and O$2p$ Kohn-Sham wavefunctions, and chooses
$P_r=P-P_d-P_p-P_{pd}-P_{dp}$. 
\item Yet, often, a Hubbard $U$ is only put on the most correlated orbitals, here
the Mn$3d$ ones. In that case, while still constructing a full pd Wannier basis, one should allow for
oxygen screening in the RPA (i.e.\ $P_r=P-P_{d}$), since those channels are blocked when
the interactions between the p and d orbitals are omitted.
Therewith, $W_r(\svek{r},\svek{r}\pr)$ of the current setup and the pd one are the same, yet their matrix-element
will differ due to the Wannier different basis.
\end{itemize}

As an outlook, we like to mention that in the described procedure to generate the effective many-body problem,
 couplings in the interaction between the low energy subspace and the
other orbital degrees of freedom are discarded -- an approximation that might prove insufficient if the subspace is chosen too restrictively.
A framework that separates the Hilbert space, while retaining the influence of self-energy effects between subspaces has been recently proposed in Ref.~\cite{ferdi_down}.

\subsection{Wannier functions and Coulomb matrix elements under pressure}
Before applying the above techniques to our compound of interest, we shall first discuss
what circumstances will influence the Wannier functions and the interaction values. 
We can distinguish three conceptually different mechanisms~: 
The choice of the orbital sub-space, the consequences of structural changes onto the localization of Wannier functions, and the changes related to the strength of screening processes.
\subsubsection{The choice of the orbital sub-space}
The construction of an effective low energy system depends very much upon the choice of the orbitals that it consists of.
In particular, the dimension of the orbital sub-space $\mathcal{H}^{low}$ has a strong impact on the localization of the Wannier functions.
Indeed, from the maximally localized Wannier function point of view, a larger subspace will allow, for a given orbital,  
for the mixing in of more Kohn-Sham orbitals. 
Therewith, the variational freedom to localize the basis functions becomes larger,
and the interaction elements bigger. See, for an example, Ref.~\cite{JPSJS.77SC.99}. 
We thus expect the interaction matrix elements of the pd-model to be larger than in the d-only model.
\subsubsection{Wannier functions under pressure}
\begin{figure*}[t!h]
  \begin{center}
    \mbox{
    \hspace{-.6cm}
\subfigure[$\,\,$ 0GPa -- B1 (NaCl)]{\scalebox{0.25}{\includegraphics[angle=-90,width=.9\textwidth]{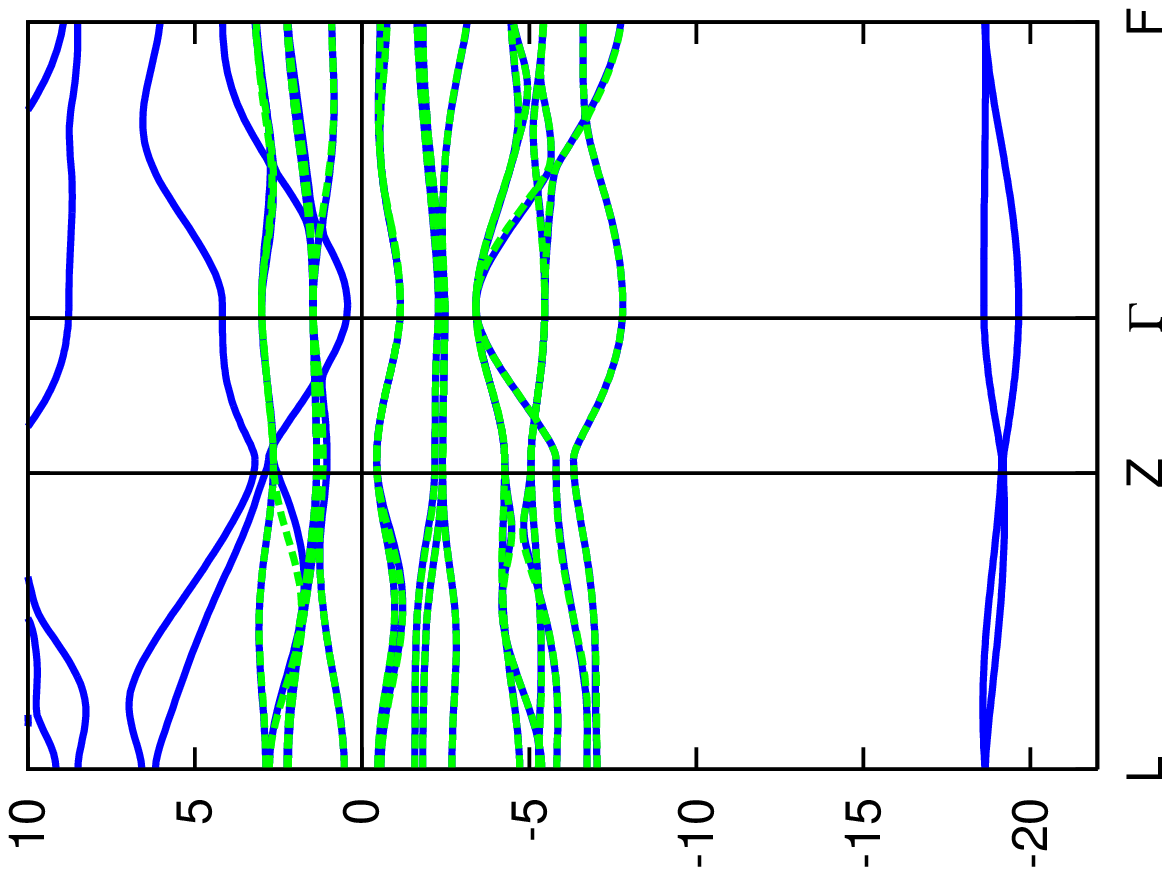}}} 
\hspace{-.6cm}
\subfigure[$\,\,$ 0GPa -- dB1]{\scalebox{0.25}{\includegraphics[angle=-90,width=.9\textwidth]{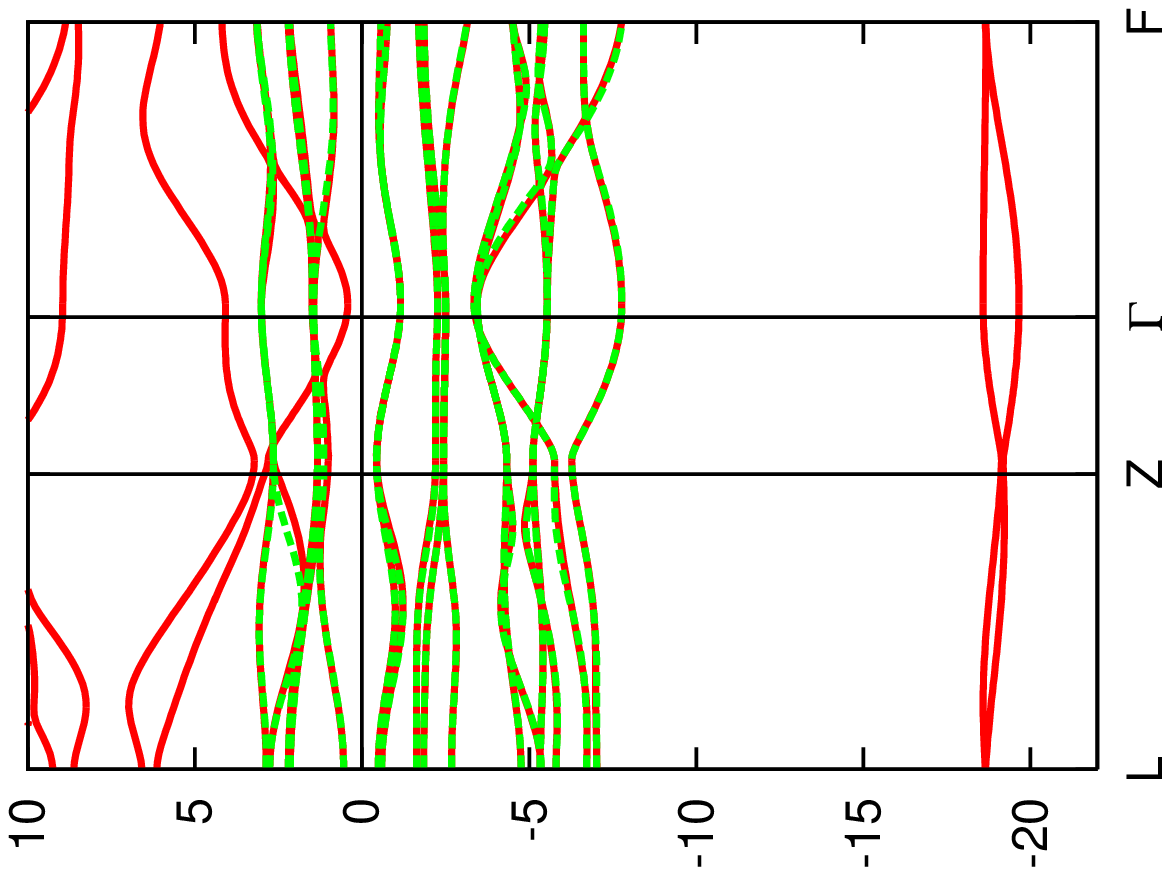}}} 
\hspace{-.6cm}
\subfigure[$\,\,$ 52.5GPa -- dB1]{\scalebox{0.25}{\includegraphics[angle=-90,width=.9\textwidth]{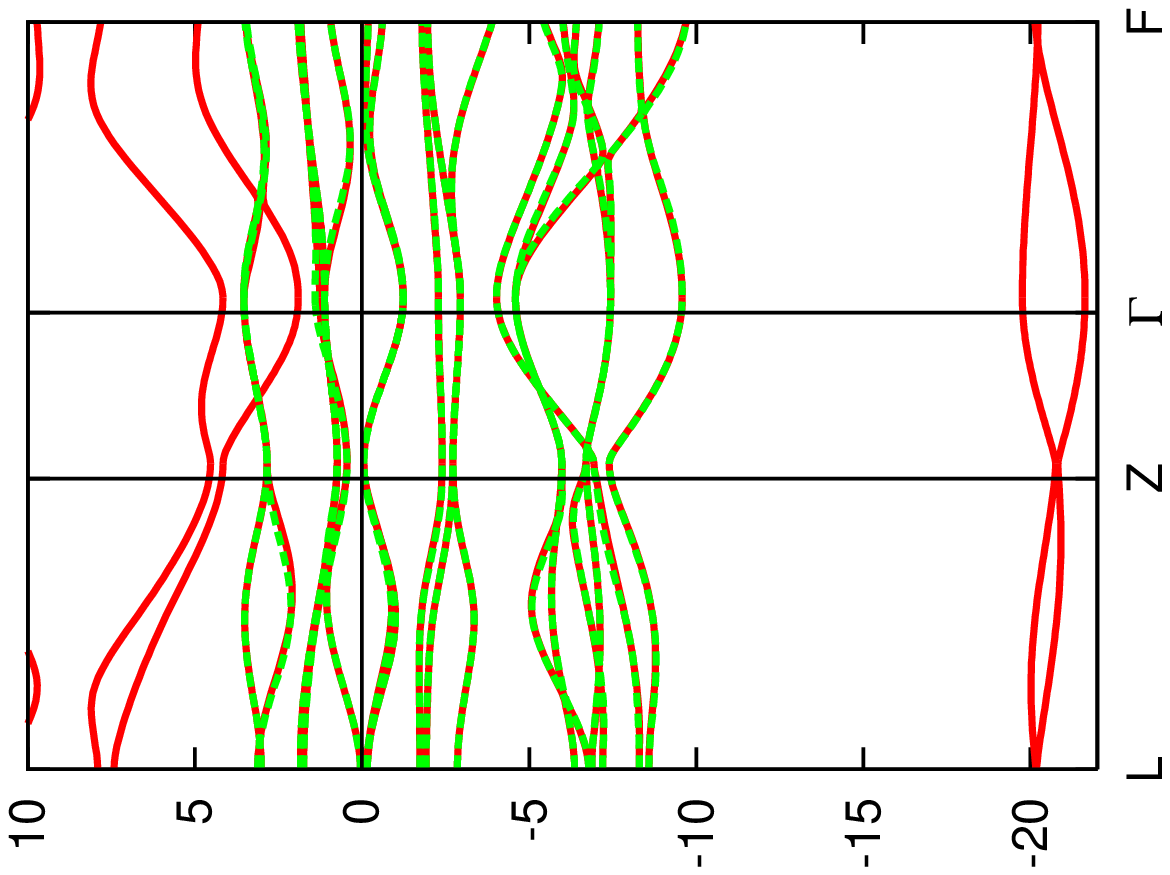}}} 
\hspace{-.6cm}
\subfigure[$\,\,$  100GPa -- dB1]{\scalebox{0.25}{\includegraphics[angle=-90,width=.9\textwidth]{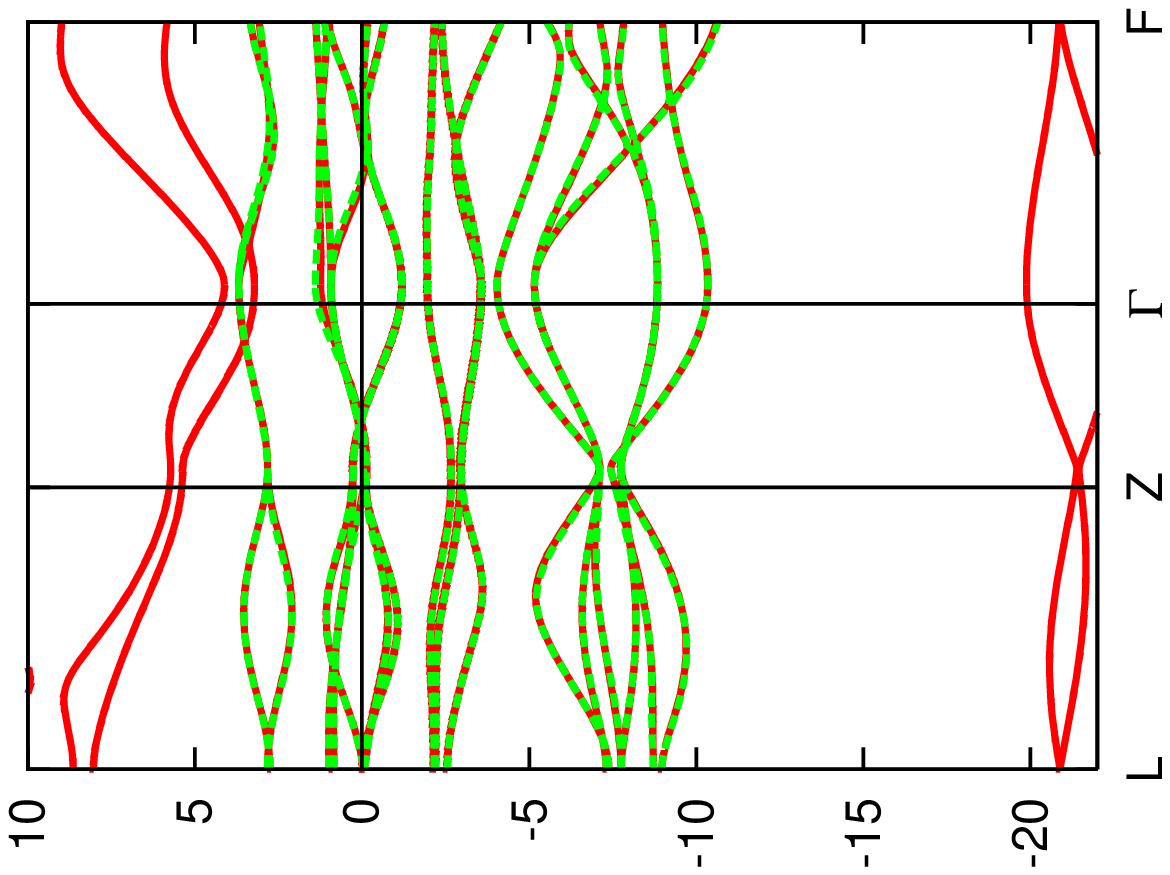}}} 
\hspace{-.6cm}
\subfigure[$\,\,$  100GPa -- B1 (NaCl)]{\scalebox{0.25}{\includegraphics[angle=-90,width=.9\textwidth]{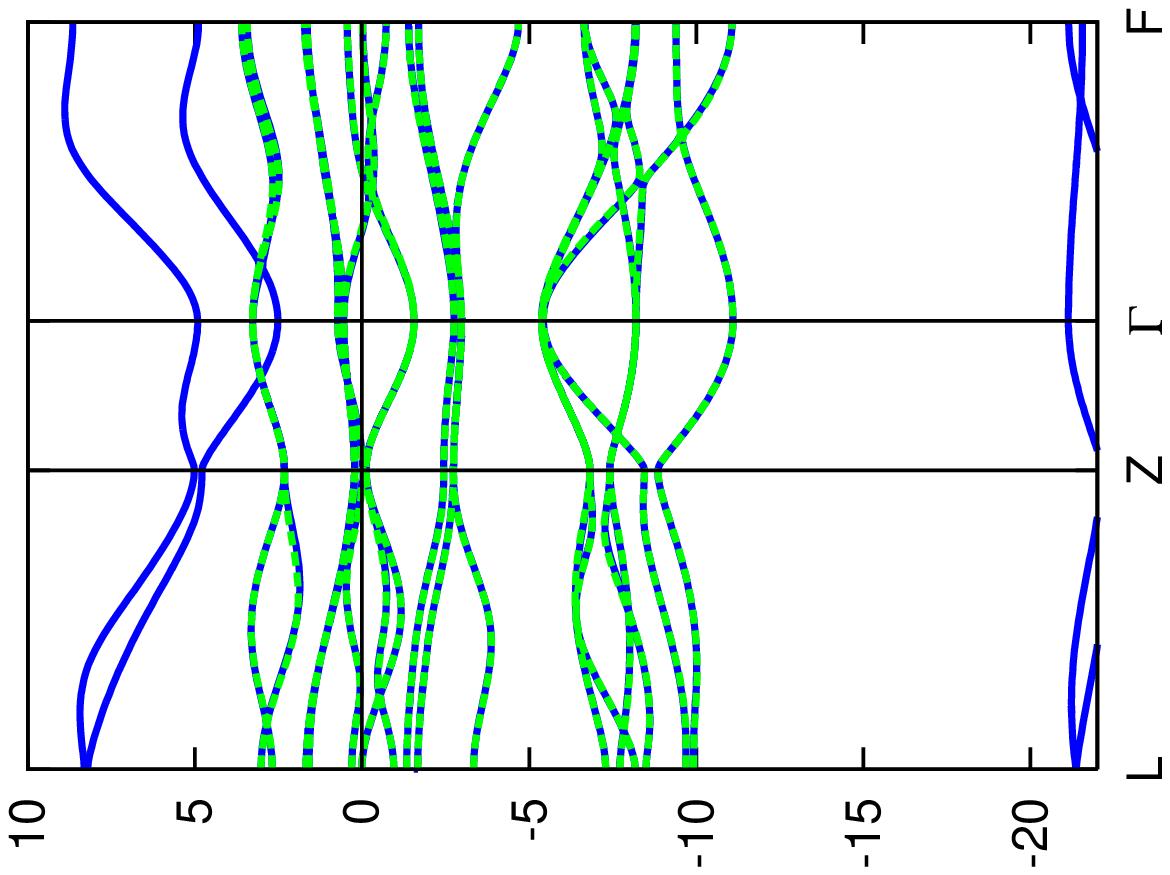}}} 
      }
    \caption{(color online) LDA band-structure of antiferromagnetic B1/dB1 MnO. Shown are the Kohn-Sham bands (solid lines) and the
    eigenvalues of the Wannier Hamiltonian (dashed). The vertical axis is energy (in eV) with the Fermi level at the origin.
     While (b) - (d) correspond to the true distorted dB1 structure at different pressures, (a) \& (d) show the band-structure
     for the artificial B1 (NaCl) structure for the indicated pressures. }
    \label{bnds_db1}
  \end{center}
\end{figure*}
External pressure, or structural changes in general, provide another impetus to alter not only the one-particle band-structure, but also the
Wannier functions and thus the bare Coulomb interaction matrix elements.

The fact that Wannier functions are not eigenfunctions even of the non-interacting problem, may result in counter-intuitive tendencies. 
In Ref.~\cite{jmt_wannier}, we established generic behaviors of maximally localized Wannier functions and interaction matrix elements under pressure. In order to put things into perspective, we shall here briefly summarize some findings of this work~:

As a model system, we consider a tight-binding parametrization of a one dimensional solid, in which case Wannier functions are 
already maximally localized if inversion symmetry is verified,\cite{PhysRevB.26.4269,PhysRevB.56.12847} and limit  the discussion to the case of a single band.
Details can be found in Ref.~\cite{jmt_wannier}.
As building blocks of the basis functions we take hydrogen-like 1s orbitals in one dimension, $\chi(x)=1/\sqrt{a_0}\exp(-\left|x\right|/a_0)$, with the Bohr radius $a_0$. The Bloch eigenfunction is given by $\psi_k(x)=A_k\sum_R e^{\im kR}\chi(x-R)$, where $A_k$ assures the normalization.
The maximally localized Wannier functions are then $\psi_R(x)=\sum_n A_n\chi(x-na-R)$, with the lattice constant $a$, while $A_n$ is a Fourier transform of $A_k$ and also incorporates the normalization of the Wannier function.
In the limit of large atomic separation ($a\gg a_0$), the Wannier functions will equal the atomic orbitals. 
Thus  
$A_n=\delta_{n,0}$ for the representative site ``0''. When pressure is applied, and the lattice constant shrinks, overlaps of the (non-orthogonal) atomic orbitals causes contributions from neighboring sites to mix into the basis functions, and the 
distribution $A_n$ broadens. This leads to an {\it increase} in the (nearest neighbor) transfer integral $t$ and a concomitant {\it increase} of the spread (as defined in~\cite{PhysRevB.56.12847}). In this sense, charge carriers become more delocalized, as expected. The tails of the Wannier functions extent over several lattice constants, before the originally exponential decay sets in. 
However, as can be shown,\cite{jmt_wannier} the coefficient $A_0$ in the above decomposition on the array of atomic orbitals, i.e.\ the strength of the  on-site orbital, grows, and thus becomes larger than one. This results in a greater probability density $\left|\psi_{R}(x)\right|^2$ around the site origins, $x-R=0$, when pressure is applied. This in turn causes a {\it larger} local Coulomb interaction matrix element $V$, \eref{eqV} .
Hence, the pressure induced delocalization is accompanied by a larger local interaction, whereas, intuitively, one might have expected the opposite. 
In our one band toy-model, since $P_r=0$ (see \eref{pol}) the screened Coulomb interaction $U$ equals the bare Coulomb interaction $V$. 
Moreover it is maximal in the maximally localized Wannier basis.\cite{jmt_wannier}
\subsubsection{Screening strength and the interaction.}
While the above changes in overlaps and hybridizations directly enter the calculation by modifying the Wannier functions, 
that influence the matrix elements of any quantity, pressure moreover enters the construction of the effective low energy system by means of the band-structure.
Indeed, according to~\eref{pol}, pressure induced modifications in the one-particle excitations 
will alter the screening strengths of possible transitions.
When bandwidths, crystal fields or bonding/antibonding splittings get enhanced upon the compression of the system, the screening
will become less effective and, therewith, the Hubbard $U$ larger. Besides this energetic effect,  evident from the denominators in \eref{pol}, there is also an effect of matrix elements when expressing the polarization in the Wannier basis.
For an example of the interplay of both effects see Ref.~\cite{jmt_wannier}.
Contrary to the preceding section, the transition energies and the matrix elements affect not the bare Coulomb interaction, but only screened quantities.
\section{Results and discussion}
\subsection{MnO in the low pressure regime~: The antiferromagnetic insulating dB1 phase}
\subsubsection{bandstructure}
\begin{figure}[t!h]
\includegraphics[angle=-90,width=.4\textwidth]{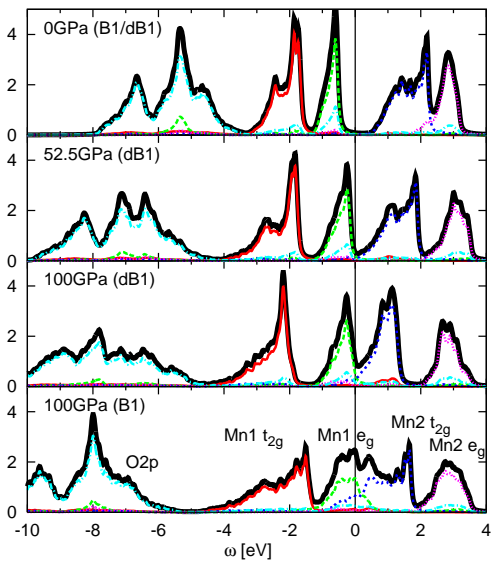}
    \caption{(Color online) Density of states (DOS) (in 1/eV) of antiferromagnetic B1/dB1 MnO for increasing pressure (top to bottom). Shown are the site-summed contributions of one spin component of the pd-model.
     The lowest panel displays fictitious B1 MnO at 100GPa at the experimental volume. Different partial DOS are depicted as indicated in the graph at the bottom. For $P=0$GPa, the DOS of the dB1 and the B1 phase do not differ significantly on the shown scale.
}
    \label{dos_db1}
\end{figure}
\fref{bnds_db1} displays the LSDA band-structure of antiferromagnetic MnO for different pressures and structures.
Explicitly marked are also the bands constructed from the Wannier orbitals
that are used for the construction of the effective ``pd'' low energy system (the bands of the effective ``d-only'' model achieve the same precision).
Using the maximally localized Wannier function formalism for entangled bands,\cite{PhysRevB.65.035109}
the choice of Hilbert space for the Mn$3d$ and O$2p$ orbitals results in a clean distinction from
the Mn 4s band that crosses the d-bands.

The corresponding LSDA densities of states (DOS) of the pd-model are shown in \fref{dos_db1}.
\begin{table}[t]	
\begin{tabular}{r||c|c|c|c} 
	P [GPa]  &$ t_{p-\t2g}$ & $\Delta_{ct}$ & $t_{p-\t2g}^2/\Delta_{ct}$ & $t_{\t2g-\t2g}$ \\ 
	\hline
	\hline
	0  (dB1)      &   0.60 & 3.5 &  0.10 & 0.21 \\
	100 (dB1)     &   0.83 & 5.5 &  0.13 & 0.20 \\
	100 $\phantom{\hbox{d}}$(B1) &   0.91 & 6.5 &  0.13 & 0.41 \\
\end{tabular}
	\caption{\label{tab1} Comparison between the direct and oxygen mediated contributions to the effective transfer element of the {\it occupied} \t2g orbitals for the B1/dB1 phase of MnO. $\Delta_{ct}=\bar{\epsilon}_{\t2g}^{occ}-\bar{\epsilon}_{2p}$ is the charge transfer energy. Shown are the {\it largest} 
	transfers for nearest neighbors of same spin. All energies in eV.}
\end{table}
\begin{figure*}[t!h!]
  \begin{center}
    \mbox{
\subfigure[ $\,\,$  on-site bare Coulomb interaction $V^{\alpha\alpha,\alpha\alpha}_{\svek{0},\svek{0}}$]{\scalebox{0.45}{\includegraphics[angle=-90,width=.9\textwidth]{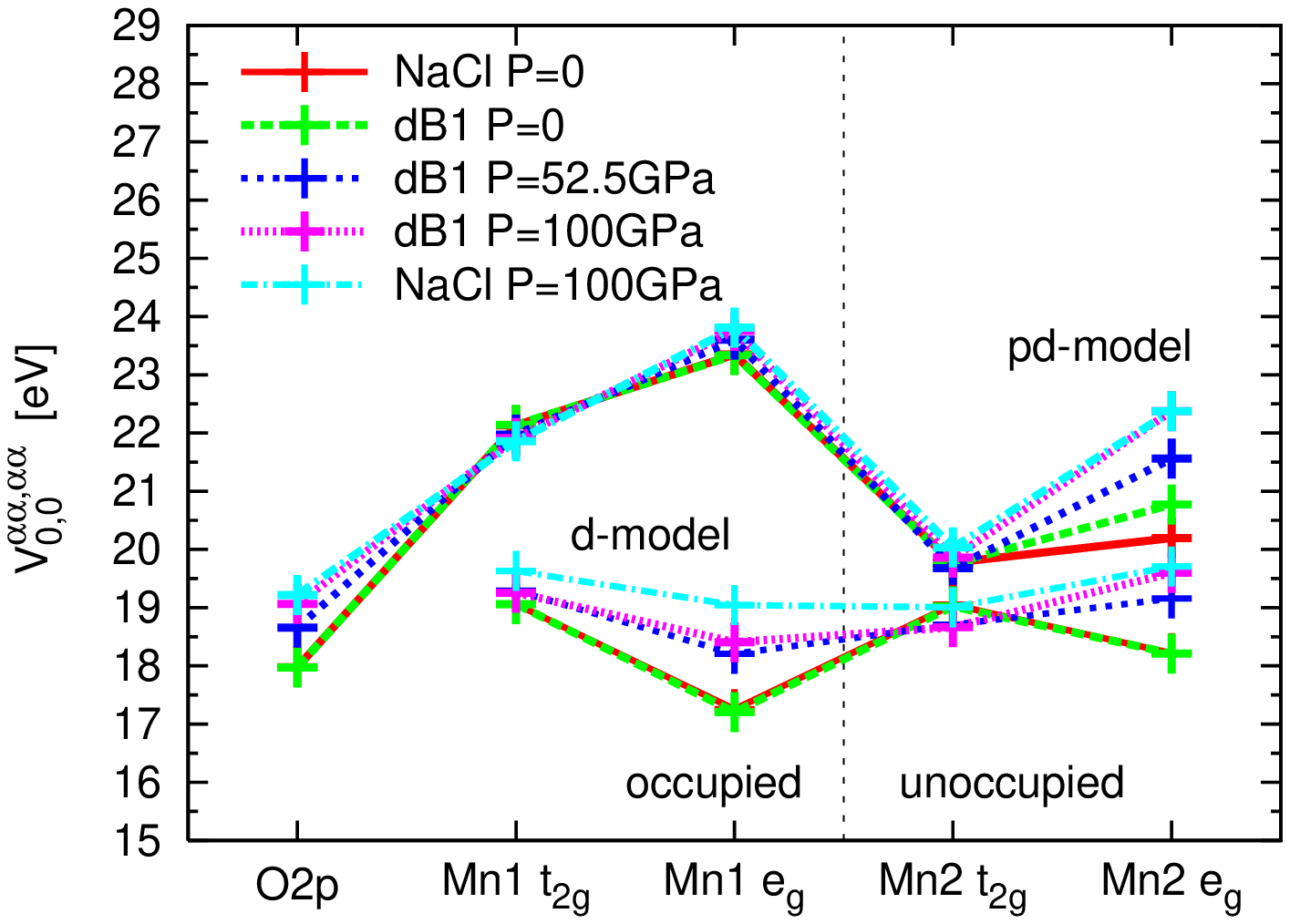}}} 
\hspace{.25cm}
\subfigure[ $\,\,$  partially screened interaction $U_{\alpha\alpha}={W_r}_{\svek{0},\svek{0}}^{\alpha\alpha,\alpha\alpha}(\omega=0)$]{\scalebox{0.45}{\includegraphics[angle=-90,width=.9\textwidth]{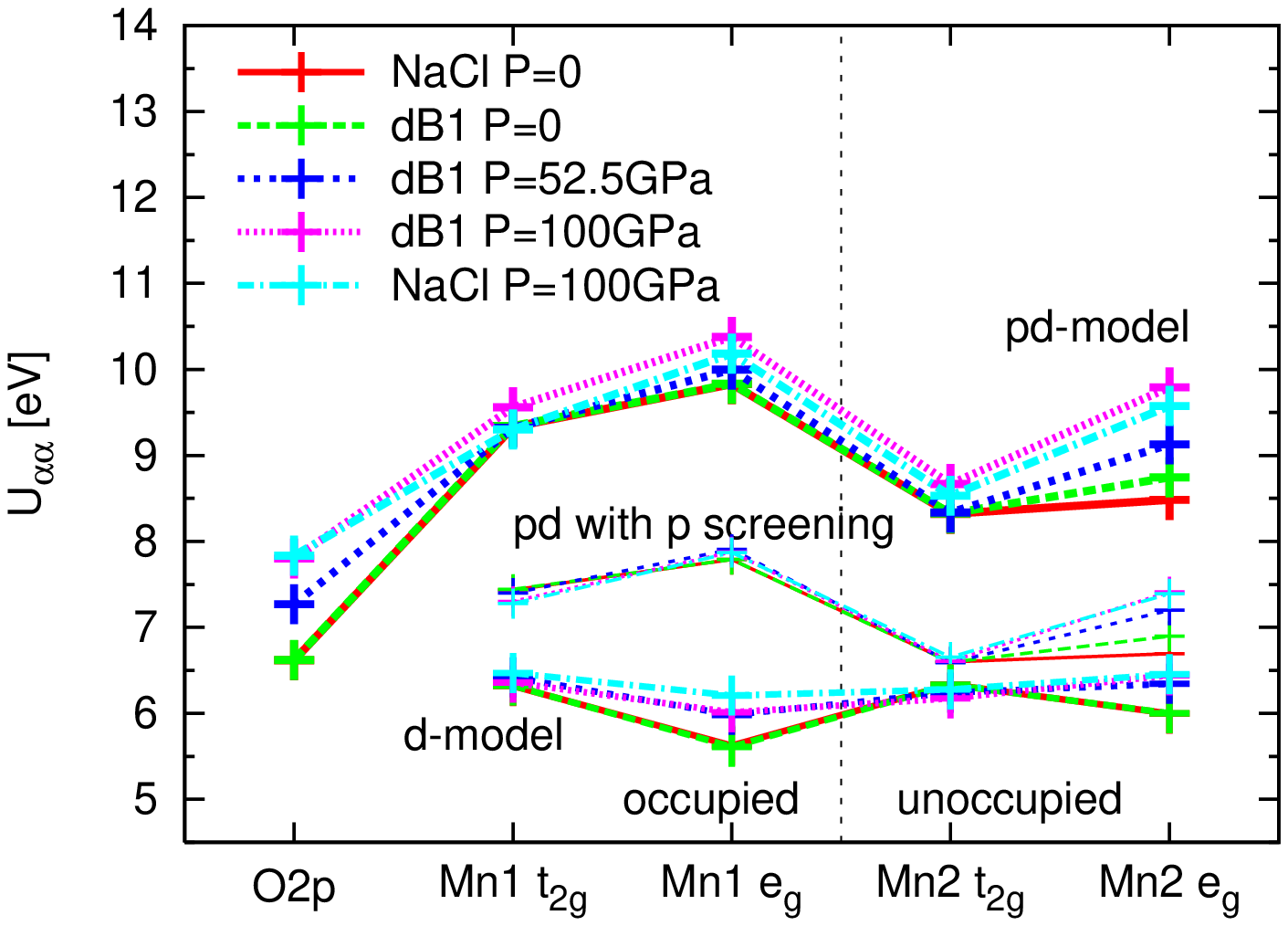}}} 
      }
    \caption{(Color online) local Coulomb interaction of the antiferromagnetic dB1 phase of MnO for different pressures. (a) diagonal elements of the local, bare interaction $V$ for the pd and d-only model, and resolved for the different orbitals $\alpha$. (b) zero frequency limit of the RPA partially screened local interaction
    $U_{\alpha\alpha}={W_r}_{\svek{0},\svek{0}}^{\alpha\alpha,\alpha\alpha}(\omega=0)$ for the pd model (with and without p screening), and the d-only model. For comparison are shown also the results of undistorted, B1 (NaCl) structured MnO.}
    \label{uv_db1}
  \end{center}
\end{figure*}
As we can see, pressure induces an increase in the crystal-field splitting, $\Delta_{cf}$, i.e.\ a larger separation between the $e_g$ and \t2g\  Kohn-Sham excitations (both the occupied and unoccupied), to an extent that the moving of the $e_g$ weight towards the Fermi level eventually causes the system to become metallic.\cite{Ohnishi} Concomitant with the increase in crystal-field, pressure results in a growing Mn--O hybridization $t_{pd}=\bra{p\svek{0}}H^{\hbox{\tiny KS}}\ket{d\svek{0}}$, where $H^{\hbox{\tiny KS}}$ is the Kohn-Sham Hamiltonian.
This effect is especially large for the $e_g$ orbitals, as is clear from the octahedral coordination. For values, see the later discussed \tref{tab1} (occupied \t2g orbitals), and \ref{tab2} ($e_g$ orbitals).
Most notably is the fact, that the partial d-bandwidths {\it do not increase with pressure}. The origin of this is identified when comparing results for the experimental dB1 phase with the undistorted B1 (NaCl) structure of the same volume. As evident from \fref{bnds_db1} (d,e), and \fref{dos_db1}, the omission of the rhombohedral distortion causes
the bandwidths to substantially increase. 
Indeed, in fictitious B1 MnO, the closing of the gap can be mainly attributed to the change in bandwidth, as was stated in Ref.~\cite{Cohen01311997,shulenburger:2009} where the distortion was neglected. 
The evolution of the bandwidths
can be quantified by looking at the effective Mn$3d$ nearest neighbor transfer elements for a given spin. The latter consists of two major contributions~: the direct hopping $t_{dd}=\bra{d\svek{0}}H^{\hbox{\tiny KS}}\ket{d\svek{R}}$ ($\svek{R}$ nearest neighbor unit-cell of $\svek{0}$), and the oxygen mediated transfer $t_{pd}^2/\Delta_{ct}$, with the hybridization
$t_{pd}$, and the charge transfer energy $\Delta_{ct}$.
The (largest) nearest (spin) neighbor transfers of the occupied \t2g orbitals, $t_{\t2g-\t2g}$, the corresponding hybridization with the oxygen $2p$ orbitals, $t_{p-\t2g}$, and the charge transfer energy $\Delta_{ct}=\bar{\epsilon}_{\t2g}^{occ}-\bar{\epsilon}_{2p}$ as measured by the band centers\footnote{In this definition, it also includes the respective contributions from the exchange splitting.}, are collected in \tref{tab1}
for ambient pressure and 100GPa, for both, the experimental dB1 and the undistorted B1 structure.

As a result of the joint increase of $t_{pd}$ and $\Delta_{ct}$, the oxygen mediated contribution to the \t2g bandwidth remains roughly of the same magnitude, irrespective of pressure and distortion.
The direct \t2g-transfer for the distorted dB1 structure is also constant with pressure. However, it is substantially structure dependent.
Indeed the nearest neighbor \t2g-\t2g transfer doubles when the distortion is neglected, causing the effective hopping to augment by 64\%,
demonstrating the critical need to include the distortion for a proper description of dB1 MnO under pressure.
As to the unoccupied \t2g orbitals, the transfers and energy differences are larger, but the same argument as above holds, for the trends are the same.
We conclude that within LDA (in contrast to previous works~\cite{Cohen01311997} (see also the recent \cite{shulenburger:2009}) relying on the undistorted structure), the gap closure under pressure is propelled by the change in crystal field and not by a broadening of band-widths.

\begin{table}[!h]	
\begin{tabular}{r||c|c|c|c} 
	P [GPa]  &$ t_{p-e_g}^{unocc (occ)}$ & $\widetilde{\Delta}_{ct}^{unocc (occ)}$ & $t_{p-e_g}^2/\widetilde{\Delta}_{ct}$ & $t_{e_g-e_g}^{unocc (occ)}$ \\
	\hline
	\hline
	0  (dB1)     								 &   1.24 (1.08)& 8.0 (4.75)&  0.19 (0.25)& 0.05 (0.06) \\
	100 (dB1)								     &   1.72 (1.50)& 10.5 (7.50)&  0.28 (0.30)& 0.13 (0.11) \\
	100 $\phantom{\hbox{d}}$(B1) &   1.82 (1.63)& 10.5 (7.50)&  0.32 (0.35)& 0.14 (0.13) \\
\end{tabular}
	\caption{\label{tab2} Comparison between the direct and oxygen mediated contributions to the effective transfer elements of the unoccupied (occupied) $e_g$ orbitals  for the B1/dB1 phase of Mn. $\widetilde{\Delta}_{ct}=\bar{\epsilon}_{e_g}-\bar{\epsilon}_{2p}$ is the charge transfer energy. Shown are the largest 
	transfers for nearest neighbors with the same spin. All energies in eV.}
\end{table}

The relevant quantities for the unoccupied (occupied) $e_g$ orbitals are compiled in Table~\ref{tab2}.
For the occupied $e_g$ orbitals the situation is opposite to the \t2g~:
While the oxygen mediated contribution to the band-width, $t_{p-e_g}^2/\widetilde{\Delta}_{ct}^{occ}$, at 100GPa depends sizeably on the structure (B1/dB1),
the direct hopping is rather insensitive in this respect.
For the unoccupied $e_g$ none of the values at 100GPa depends much on the structure, as can be inferred also from the similar shapes of the density of states.
Yet, overall the effective bandwidths of the dB1 phase are always a little smaller than for the undistorted phase.
Thus, the rhombohedral distortion is effectively reducing all partial d-bandwidths. 
\subsubsection{The Coulomb interaction}
\paragraph{The bare interaction.}
As explained in a model context in Ref.~\cite{jmt_wannier}, changes in the orbital overlaps and hybridizations are expected to have an impact on the bare Coulomb interaction matrix $V$ of \eref{eqV}. The on-site elements of the latter for antiferromagnetic dB1 MnO are shown in \fref{uv_db1}(a), both for a pd and a d-only Wannier basis set, and for different pressures.

Owing to the distortion axis, the degeneracy properties of the Mn$3d$ orbitals do not change with pressure.
Moreover, in the current case, the maximally localized Wannier functions retain the degeneracy of the LMTOs, and while the LDA wavefunctions were merely the initial guesses for the maximally localized Wannier functions, we keep the labeling of
``\t2g'', ``$e_g$'', ``$d_{xy}$'' and alike. 
In some cases, especially for entangled bands, the localization procedure might cause the resulting Wannier functions to acquire a different symmetry. 

In the case of the pd-model, the matrix elements are sensitive to pressure only for the O$2p$ and the unoccupied $e_g$ orbitals.
There, the changes in the interaction are quite significant and reach up to 10\%.
This is a natural consequence of our above remarks~: It is precisely those orbitals whose bondings changes the most under pressure, namely
the hybridizations between them increase considerably. As summarized in the Method section, and discussed in detail in Ref.~\cite{jmt_wannier}, this allows the respective Wannier functions to accumulate more weight at
small distance, while an overall increase in spread accounts for the gain in hybridization.
This is also the reason why, for the pd-model, the $e_g$ elements are larger than the $t_2g$ ones~: the larger hybridizations of the $e_g$ orbitals with the oxygen $2p$ ones, allow the corresponding maximally localized Wannier functions to reduce their extension.

As anticipated, the elements of the d-model are lower than for the pd case, since the Wannier functions are more extended.
Having restrained the variational freedom to the d orbitals, their matrix elements acquire a larger dependence on pressure than in the pd case.
This is a result of the fact that the intra-d hoppings, especially the $e_g$ ones, must make up for the dispersion caused by oxygen mediated transfers that are not present
in the smaller basis.
 
 \begin{table}[!t]	
\begin{tabular}{r||c|c}
	P [GPa]  & $U_{pd}$ [eV]& $U_{dd}^{nn}$ [eV]\\
	\hline
	\hline
0.0 (dB1) & 2.25 & 1.51\\
52.5 (dB1)& 2.44& 1.69\\
100.0 (dB1)& 2.65 & 1.96\\
100.0 $\phantom{\hbox{d}}$(B1)& 2.61 & 1.74 \\
	\hline
	\hline
96.5 $\phantom{\hbox{d}}$(B8)& 1.66 &1.11\\
120.0 $\phantom{\hbox{d}}$(B8)& 1.86 & 1.26\\
160.0 $\phantom{\hbox{d}}$(B8)& 1.98 & 1.42
\end{tabular}
	\caption{\label{tab3} Averaged nearest neighbor inter-atomic interactions $U_{pd}={W_r}^{pp,dd}_{\svek{0},\svek{0}}(\omega=0)$ and $U_{dd}^{nn}={W_r}^{d_1d_1,d_2d_2}_{\svek{0},\svek{0}}(\omega=0)$  (where $d_{1/2}$ is a short hand for the d-orbitals of the 1./2. Mn atom in the cell) for the pd-model of the B1/dB1 phase (top) and the B8 phase (bottom) for different pressures. 
	}
\end{table}
 \begin{figure*}[t!h]
  \begin{center}
    \mbox{
    \hspace{-.25cm}
\subfigure[$\,\,$  96.5GPa -- B8]{\scalebox{0.29}{\includegraphics[angle=-90,width=\textwidth]{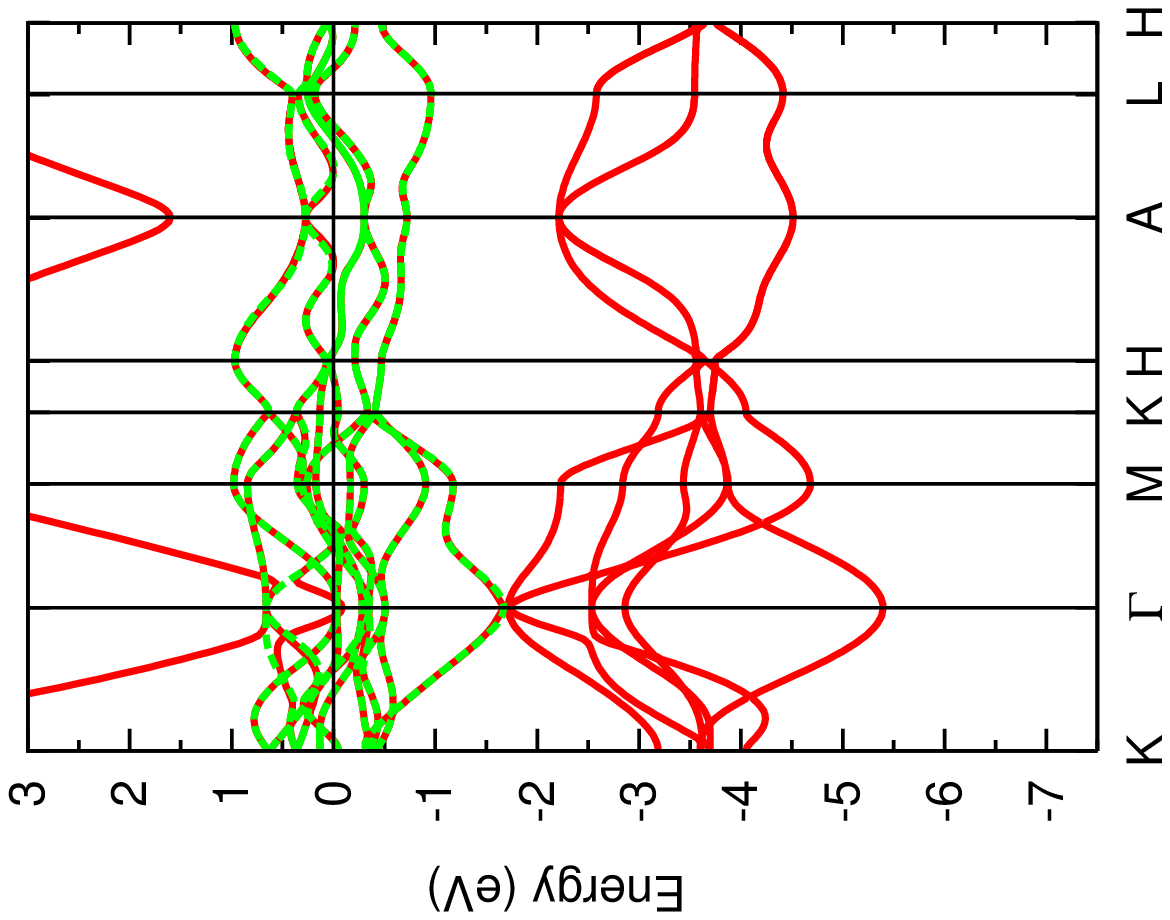}}} 
\subfigure[$\,\,$ 120GPa -- B8]{\scalebox{0.29}{\includegraphics[angle=-90,width=\textwidth]{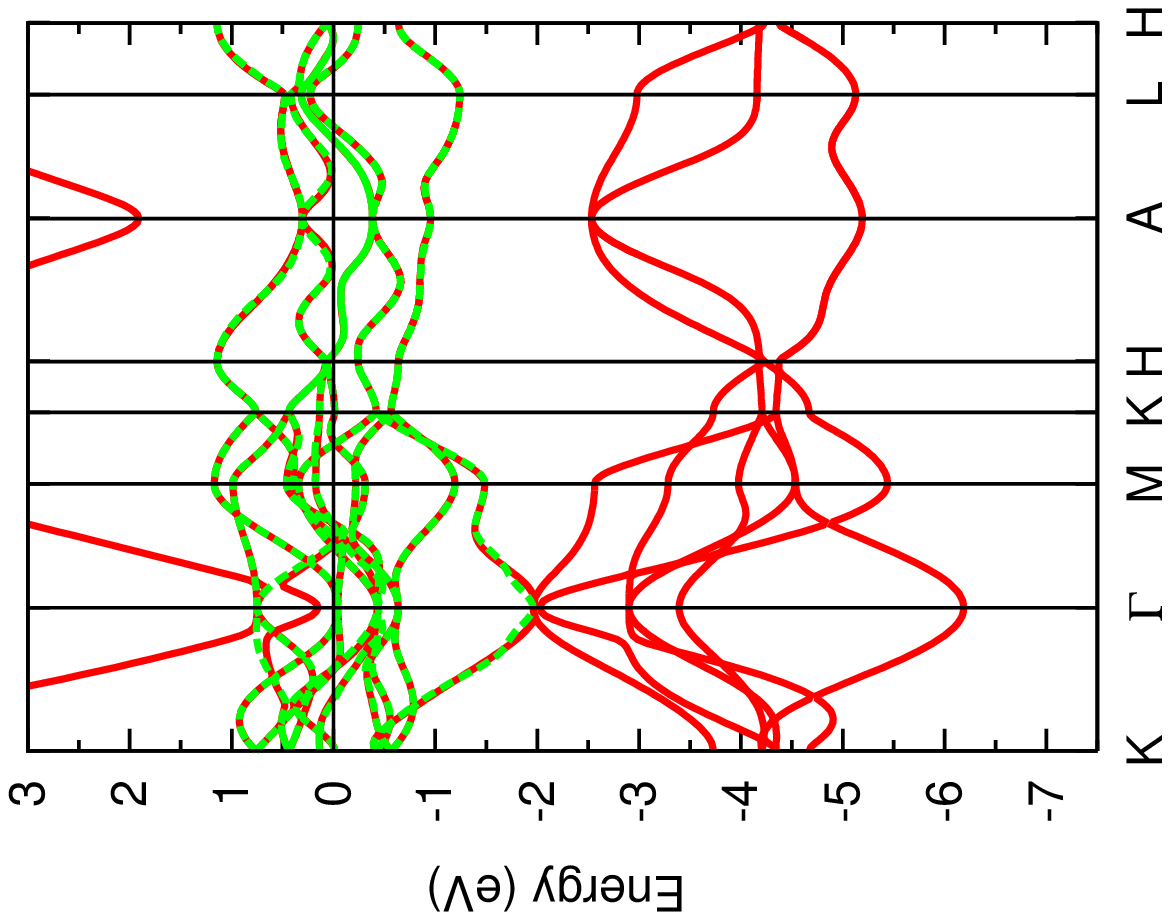}}} 
\subfigure[$\,\,$ 160GPa -- B8]{\scalebox{0.29}{\includegraphics[angle=-90,width=\textwidth]{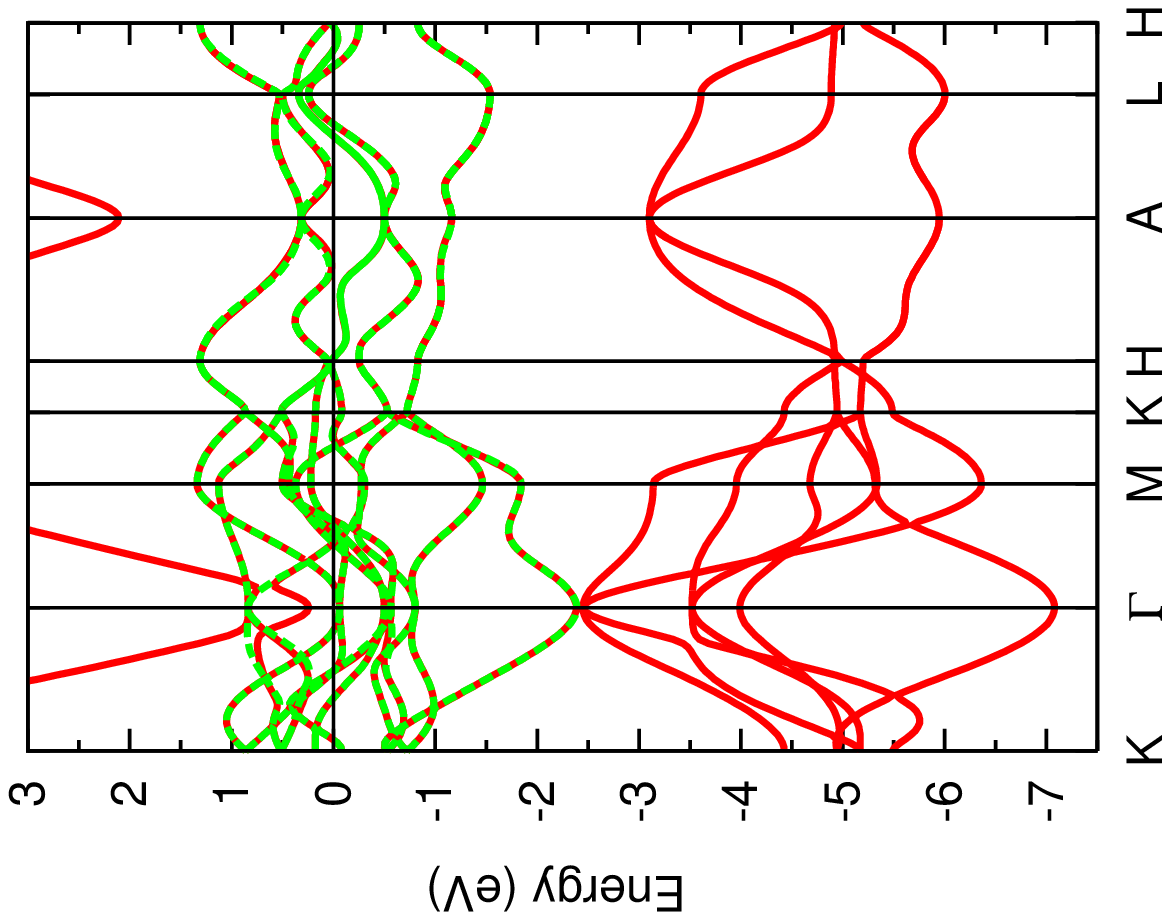}}} 
      }
    \caption{(color online) LSDA band-structure of paramagnetic MnO in the B8 high pressure phase. Shown are the Kohn-Sham bands (solid) and the
    eigenvalues of the Wannier Hamiltonian (dashed).  The vertical axis is energy (in eV) with the Fermi level at the origin.
  }
    \label{bnds_b8}
  \end{center}
\end{figure*}
\paragraph{The partially screened interaction.} 

While the bare Coulomb interaction was mainly a mean to analyze the pressure dependence of the Wannier functions, the quantity that is needed for the formulation of the effective models is the partially screened Coulomb interaction. In \fref{uv_db1} (b) we show, for the same pressures as before, the Hubbard $U$ for the three different types of models, as introduced in the Method's section~: the full pd-model, the pd-model with pd-screening (for a treatment in which only the d-orbitals will be supplemented by a local interaction), and the d-only model.
The magnitudes of the corresponding interactions decrease in the given order~: The interactions in the pd-model are smaller when the p-screening is present, and the interactions in the d-only model are yet smaller because of the larger extension of the Wannier functions.
The pressure dependence of the elements has the same tendencies as the bare Coulomb interactions. Yet, we note that the relative difference,  $(U^{100GPa}-U^{0GPa})/U^{0GPa}$,  is larger for the Hubbard $U$ than for $V$, the pressure influence thus stronger. This is understood from a decrease in screening strength owing to the larger separation in energy between occupied and unoccupied states as pressure increases (see e.g.\ Ref.~\cite{jmt_wannier} for a discussion), as apparent from the band-structure, \fref{dos_db1}. On an absolute scale, and for the given pressure range, the Hubbard $U$ changes by up to 1~eV for both the Mn$3d$ and the O$2p$ orbitals -- a non-negligible effect given the aforementioned modifications in the magnitude of the transfer integrals.

For the sake of completeness and to put the values further into perspective, Table~\ref{tab3} (upper part) contains the nearest neighbor partially screened interactions
$U_{pd}={W_r}^{pp,dd}_{\svek{0},\svek{0}}(\omega=0)$ and $U_{dd}^{nn}={W_r}^{d_1d_1,d_2d_2}_{\svek{0},\svek{0}}(\omega=0)$ 
 (where $d_{1/2}$ is a short hand for the d-orbitals of the 1./2. Mn atom in the cell)
within the pd-model and averaged over the respective orbital subset.
Both elements are roughly a factor of four smaller than the intra-atomic interactions.
\begin{widetext}
\begin{center}
\begin{table}[!h]	
\begin{tabular}{r||c|c|c|c|c} 
	$\phantom{9}$0 (100) GPa & $d_{xy}$ & $d_{xz}$ & $d_{z^2}$ & $d_{yz}$ & $d_{x^2-y^2}$ \\
	\hline
	\hline
 $d_{xy}$ & -- &  0.61 (0.62) &  0.66 (0.68) &  0.61 (0.62) &  0.38  (0.37) \\  
 $d_{xz}$ & 0.61  (0.62) &  -- &  0.46 (0.45) &  0.61 (0.62) &  0.59  (0.60) \\  
 $d_{z^2}$ & 0.66  (0.68) &  0.46 (0.45) &  -- &  0.45 (0.45) &  0.58  (0.65) \\
 $d_{yz}$  & 0.61  (0.62) &  0.61 (0.62) &  0.45 (0.45) &  -- &  0.59  (0.60) \\  
 $d_{x^2-y^2}$ & 0.38  (0.37) &  0.59 (0.60) &  0.58 (0.65) &  0.59 (0.60) &  -- 
\end{tabular}
\caption{\label{tab4} Hund's coupling $J={W_r}_{\svek{0},\svek{0}}^{\alpha\beta,\beta\alpha}(\omega=0,\alpha\ne\beta)$ of antiferromagnetic dB1 MnO and the d-only model for the extreme pressures 0 and 100GPa.} 
\end{table}
\end{center}
\end{widetext}
As a function of pressure, both $U_{pd}$ and $U_{dd}^{nn}$
increase. As was true for the diagonal elements of the Hubbard U, the pressure dependence of these off-diagonal interactions is a feature mostly intrinsic to the Wannier functions and not the screening strength.

The Hund's coupling, $J={W_r}_{\svek{0},\svek{0}}^{\alpha\beta,\beta\alpha}(\omega=0,\alpha\ne\beta)$, is compiled below in Table~\ref{tab4}. While being largely anisotropic with respect to the orbital (for the Mn d orbitals, values range from 0.4~eV to 0.7~eV at ambient pressure), it does not depend significantly on pressure.
%

%
\subsection{MnO at high pressure~: the paramagnetic B8 phase}
\setcounter{paragraph}{0}
\subsubsection{The band-structure}

The B8 phase of MnO is of NiAs (B8) structure. As discussed in Ref.~\cite{yoo:115502}, the B8 and dB1 structure are rather similar, and a transition from the high pressure dB1 to the B8 phase requires a moving of oxygen atoms from ``distorted octahedral'' to ``perfect trigonal'' positions.

\begin{figure}[!h]
  \begin{center}
\includegraphics[angle=-90,width=.425\textwidth]{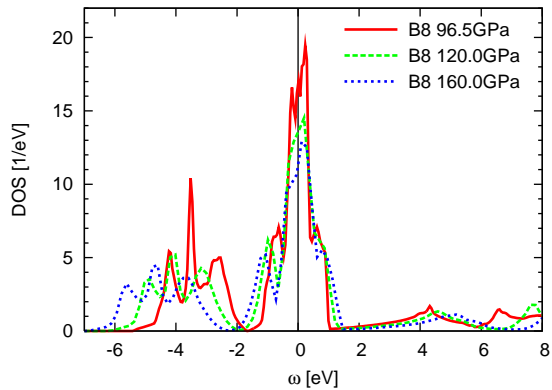}
    \caption{(Color online) LDA density of states (DOS) of the paramagnetic B8 phase of MnO for various pressures. The Fermi level corresponds to $\omega=0$. 
    }
    \label{dos_b8}
  \end{center}
\end{figure}

Yet, in the pressure dependence, there is an important difference~: Augmenting pressure in the dB1 phase leads to an increase of the 
trigonal angle, 
and thus to anisotropic changes. 
This results in an almost pressure-insensitive bandwidths, as discussed above. 
In the B8 phase, the c/a ratio basically remains unchanged, and it is thus only the uniform modification of the lattice constant that accounts for the change in volume.
We note that the different coordination causes the d-orbitals to split into two doubly degenerate $e_g^{1,2}$ and a single $a_g$ orbitals (see e.g.\ Ref.~\cite{1367-2630-9-7-235}).
\fref{bnds_b8} shows the band-structure of B8 MnO for various pressures, along with the bands in the Wannier gauge for the d-model.
The corresponding DOS is displayed in
\fref{dos_b8}. 
The lowest indicated pressure is in fact smaller than the largest value we had chosen for the dB1 phase. This is justified by the fact that both phases are 
coexisting over a finite pressure range, as evidenced from optics~\cite{Kobayashi01,mita:100101} and x-ray experiments.\cite{yoo:115502}

Of course, the LDA yields a metal for all pressures. Since the Mott transition within the paramagnetic B8 phase only takes place at finite temperatures, this is congruent with experiments. However, we used again experimental structures and volumes at ambient temperature,\cite{yoo:115502} since our aim is the construction of a model for the metal-insulator transition within the B8 phase. 
Continuing our remarks from the Methods section, this model construction is approximate. 
While it is perfectly admissible that the one-particle 
dispersion is metallic, one might object that the pd screening that we include in the determination of the Hubbard $U$ could be overestimated since the exact Green's function
will have excitations with a larger separation because of the insulating nature of the system.

In the LDA band-structure, \fref{bnds_b8},
the d-bandwidth becomes slightly larger with increasing pressure. As mentioned in Ref.~\cite{kunes_mno} this restricted increase is a result of the fact that the d-bandwidth is mainly constituted by the hybridization contribution of the O$2p$ orbitals. Yet, while $t_{pd}$ increases with pressure, so does the charge transfer energy $\Delta_{ct}$, and thus the effective hopping, $t_{pd}^2/\Delta_{ct}$, does not increase much (cf. our discussion of the dB1 phase.).
The preponderant effect of pressure on the band-structure is a moving away from the Fermi surface of the occupied O$2p$ and the unoccupied Mn $4s$, $4p$ and $4d$ orbitals, owing to greater hybridization splittings. 
\subsubsection{The interaction}
\begin{figure*}[!t!h!]
  \begin{center}
\subfigure[ $\,\,$ on-site bare Coulomb interaction $V^{\alpha\alpha,\alpha\alpha}_{\svek{0},\svek{0}}$]{\scalebox{0.45}{\includegraphics[angle=-90,width=.9\textwidth]{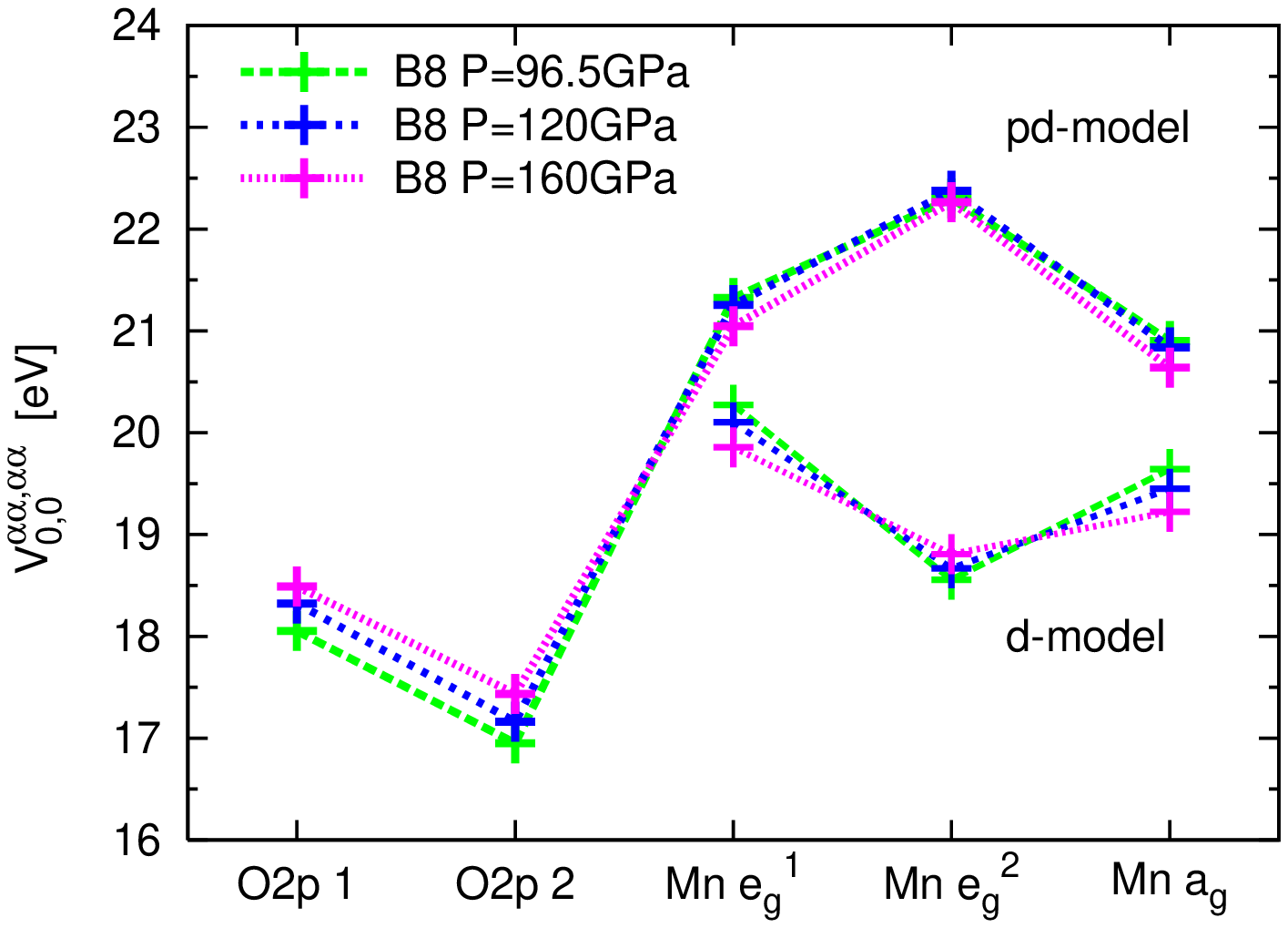}}} 
\hspace{.25cm}
\subfigure[ $\,\,$ partially screened interaction $U_{\alpha\alpha}={W_r}_{\svek{0},\svek{0}}^{\alpha\alpha,\alpha\alpha}(\omega=0)$]{\scalebox{0.45}{\includegraphics[angle=-90,width=.9\textwidth]{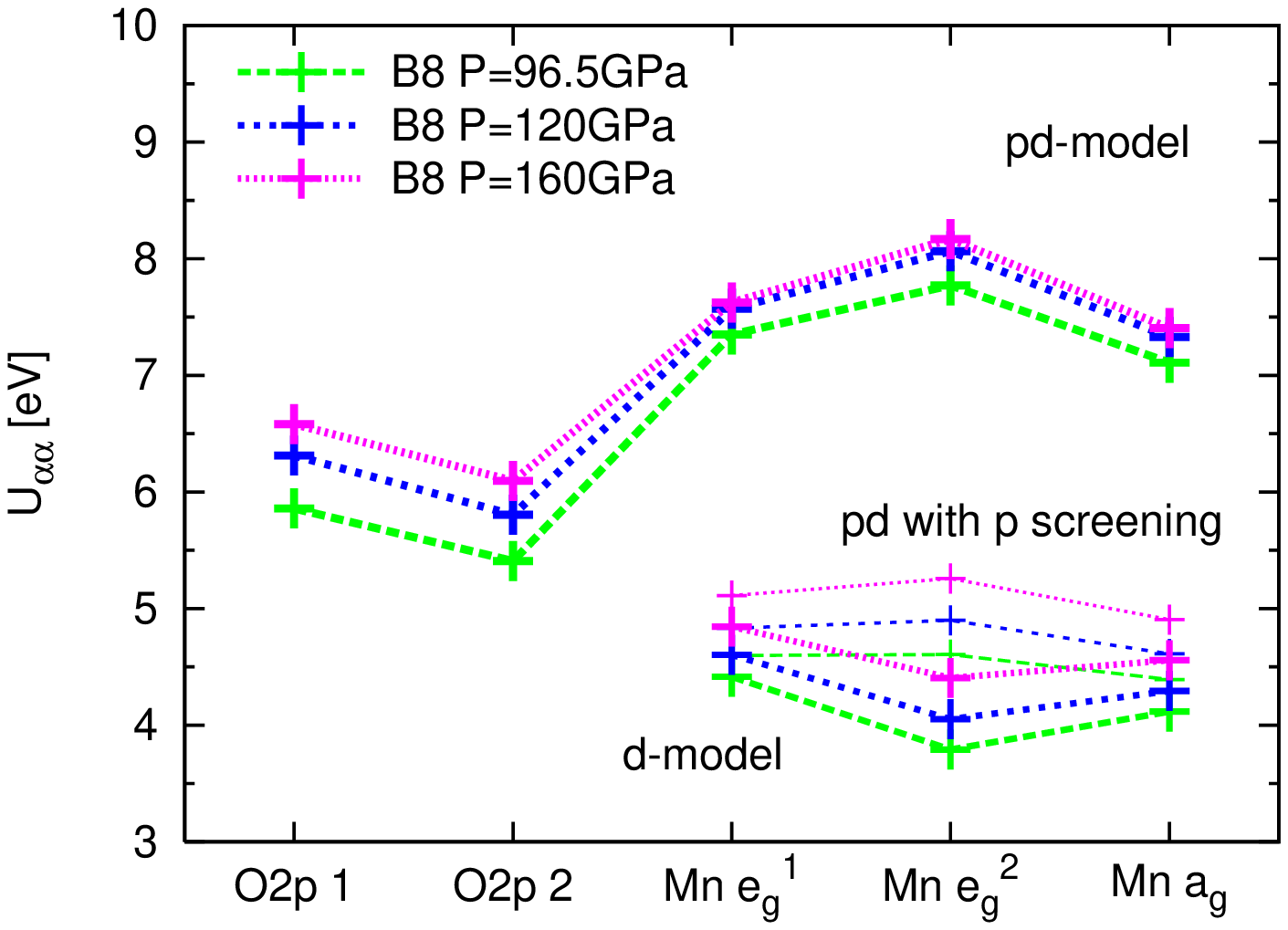}}} 
    \caption{(Color online) local Coulomb interaction of the paramagnetic B8 phase of MnO for different pressures. (a) diagonal elements of the local, bare interaction $V^{\alpha\alpha,\alpha\alpha}_{\svek{0},\svek{0}}$ for the pd and d-only model, and resolved for the different orbitals $\alpha$. (b) zero frequency limit of the RPA partially screened local interaction
    $U_{\alpha\alpha}={W_r}_{\svek{0},\svek{0}}^{\alpha\alpha,\alpha\alpha}(\omega=0)$ for the pd model (with and without p screening), and the d-only model. }
    \label{uv_b8}
  \end{center}
\end{figure*}
Given the above changes in bandwidths, the bare Coulomb interaction is expected to be roughly pressure independent.
As can be inferred from \fref{uv_b8} (a), which displays this quantity for both the pd and the d-only model, this is found indeed to be the case.
The absolute values are comparable in size to those of the dB1 phase.

The pressure dependence is decisively different for the partially screened interaction --the Hubbard U-- shown in \fref{uv_b8}, as above, again for the three different types of models.
As pressure augments, the changes induced to the bands-structure lead to a decrease in screening strength, resulting in an almost isotropic increase of the effective interaction of up to about 0.1 eV/10GPa for the pd-model.

The average nearest neighbor interaction elements $U_{pd}$ and $U_{dd}^{nn}$ (for definitions see above) for the pd-model
are compiled in \tref{tab3}. The trend with pressure is the same as for the intra-atomic elements, i.e.\ they grow with pressure. However, like in the dB1 phase, they are smaller by a factor of about four with respect to the intra-atomic elements.
While the pressure dependence of the diagonal terms of the screened interaction could be attributed almost exclusively to changes in the screening,
it is notable that the off-diagonal elements of already the bare Coulomb matrix elements change by almost 10\%  from the lowest to the highest pressure considered.

What had been said about the Hund's coupling of the dB1 phase is also true here, as seen from Table~\ref{tab5}. Being less susceptible to screening processes, it does not show a significant pressure dependence. However, it is, again, sensitive to the combination of orbitals involved, and values can vary by as much as 65\%.
Since the physics of the high spin to low spin  transition~\cite{kunes_mno} within the B8 phase is sensitive not to the diagonal elements of the interactions (which change considerably), but mostly to changes in J~\footnote{J. Kune\v{s}, private communication.} (that are small), arguments made about the transition mechanism~\cite{kunes_mno} may not be affected when using the pressure dependent matrix elements instead of constant ones. Yet, spectra will change on a quantitative level.

\begin{widetext}
\begin{center}
\begin{table}[!h]	
\begin{tabular}{r||c|c|c|c|c} 
	96 (160) GPa  & $d_{xy}$ & $d_{xz}$ & $d_{z^2}$ & $d_{yz}$ & $d_{x^2-y^2}$ \\
	\hline
	\hline
$d_{xy}$ & --         &  0.49   (0.51) &  0.70   (0.69) & 0.61   (0.61) &   0.58  (0.54) \\
$d_{xz}$&0.49  (0.51) &  -- &             0.48   (0.46) & 0.63   (0.65) &   0.62 (0.63) \\
$d_{z^2}$&0.70  (0.69) &  0.48   (0.46) &  -- &           0.48   (0.49) &   0.69  (0.69) \\
$d_{yz}$&0.61  (0.61) &  0.63   (0.65) &  0.48   (0.49) & -- &              0.49  (0.51) \\
$d_{x^2-y^2}$&0.58  (0.54) &  0.62   (0.63) &  0.69   (0.69) & 0.49   (0.51) &   -- \\
\end{tabular}
	\caption{\label{tab5} Hund's coupling $J={W_r}_{\svek{0},\svek{0}}^{\alpha\beta,\beta\alpha}(\omega=0,\alpha\ne\beta)$ of paramagnetic B8 MnO for the d-only model and the  pressures 96 and 160GPa.  
	} 
\end{table}
\end{center}
\end{widetext}

\section{Summary, Conclusions and Perspectives}
In conclusion, we have studied the impact of external pressure onto the construction of effective low energy many-body models for the realistic example of manganese monoxide. 
This compound is particularly challenging for a first principle description, since energies, such as the transfer integrals, the Hubbard $U$ and the charge transfer energies are of the same magnitude.
On the level of band-structures, we concluded that the pressure induced vanishing of the gap in the antiferromagnetic phase of MnO is driven 
by the crystal field rather than by an increase of band-widths, the latter of which is found to be limited by the growing rhombohedral distortion.
The high pressure paramagnetic phase, on the contrary, exhibits the usual pressure caused increase of hybridizations and bandwidths.
Using the constraint RPA approach within setups of maximally localized Wannier functions, we further found that, for a quantitative description, the pressure dependence of the Hubbard $U$ can not be neglected. While the pressure sensitivity in the low pressure antiferromagnetic phase of MnO is dominated by compression induced changes of the bare Coulomb interaction, the effective interaction of the high pressure paramagnetic phase is mostly determined by modifications in the band-structure that affect the screening properties of the system.
Our {\it ab initio} constructed low energy models can provide a quantitative starting point for the application of sophisticated many-body methods. 

\section*{Acknowledgments}
We gratefully acknowledge discussions with and comments by R. Sakuma, J. Kune\v{s}, K. Terakura and  R. Cohen.
This work was in part supported by
the G-COE
program of MEXT(G-03) and 
 the Next Generation Supercomputer Project, Nanoscience Program
from MEXT, Japan.


\end{document}